\documentclass[twocolumn,showpacs,preprintnumbers,amsmath,amssymb,nofootinbib]{revtex4}
\usepackage{dcolumn}
\usepackage{bm}
\usepackage{tabularx}
\usepackage{array}
\usepackage{graphics}
\usepackage{graphicx}
\usepackage{psfrag}
\usepackage{epsfig}
\usepackage{amsmath}
\usepackage{amssymb}
\usepackage{verbatim}
\usepackage{color}
\usepackage{psfrag}

\usepackage{verbatim}

\newcommand{\msbar}{{\overline{\rm MS}}}

\newcommand{\gpi}{g_{D^\ast D\pi}}

\newcommand{\nf}{N_{\rm f}}
\newcommand{\bea}{\begin{eqnarray}}
\newcommand{\eea}{\end{eqnarray}}
\newcommand{\beq}{\begin{equation}}
\newcommand{\eeq}{\end{equation}}
\newcommand{\gev}{{\rm GeV}}
\newcommand{\mev}{{\rm MeV}}
\newcommand{\kev}{{\rm keV}}

\newcommand{\pdir}{p\kern -5.2pt\raise 0.2ex\hbox {/}}
\newcommand{\vdir}{v\kern -5.75pt\raise 0.15ex\hbox {/}}
\newcommand{\kdir}{k\kern -5.75pt\raise 0.15ex\hbox {/}}
\newcommand{\epsdir}{\epsilon\kern -5.0pt\raise 0.15ex\hbox {/}}
\newcommand{\bvdir}{\bar{v}\kern -5.75pt\raise 0.15ex\hbox {/}}
\newcommand{\Ddir}{D\kern -7.75pt\raise 0.20ex\hbox {/}}
\newcommand{\Adir}{A\kern -7.75pt\raise 0.20ex\hbox {/}}
\newcommand{\ldir}{l\kern -5.0pt\raise 0.2ex\hbox{/}}
\newcommand{\varepsdir}{\varepsilon\kern -5.5pt\raise 0.15ex\hbox{/}}

\newcommand{\nn}{\nonumber}
\makeatother

\begin{document}
\preprint{\tt LPT Orsay 12-105}
\vspace*{22mm}
\title{Theoretical estimate of the $D^\ast \to D\pi$ decay rate}

\author{Damir Be\v{c}irevi\'c}
 \email{Damir.Becirevic@th.u-psud.fr}
\affiliation{%
Laboratoire de Physique Th\'eorique (B\^at 210), Universit\'e Paris Sud,  
Centre d'Orsay, 91405 Orsay-Cedex, France}

\author{Francesco Sanfilippo} 
\email{Francesco.Sanfilippo@th.u-psud.fr}
\affiliation{%
Laboratoire de Physique Th\'eorique (B\^at 210), Universit\'e Paris Sud,  
Centre d'Orsay, 91405 Orsay-Cedex, France}

\date{\today}

\begin{abstract}
We present the results of our lattice QCD study of the $\gpi$ coupling, relevant to the $D^\ast \to D\pi$ decay. 
Our computation is made on the gauge field configurations that include $\nf=2$ dynamical light quarks by using the twisted mass QCD action. 
From the results obtained at four different lattice spacings we were able to take for the first time the continuum limit of this quantity computed on the lattice. Our final value, 
$g_c=0.53(3)(3)$, leads to $\Gamma(D^{\ast +}\to \overline D^0 \pi^+)=\left(50\pm 5\pm 6\right)$~keV, and is in good agreement with the experimental 
results for the width of the charged $D^\ast$-meson. 
\end{abstract}

\pacs{12.39.Fe, 12.39.Hg, 13.20.-v, 11.15.Ha.}
\maketitle

\section{\label{Introduction}Introduction}
The result of the first experimental measurement of the width of the charged vector $D$-meson, $\Gamma(D^{\ast +})=96\pm 22$~keV~\cite{cleo},  
was a surprise because many theoretical predictions of the relevant hadronic coupling, $\gpi$, suggested much smaller (by about a factor of $3\div 4$) value for $\Gamma(D^{\ast +})$.  An important novelty this year is that the value of $\Gamma(D^{\ast +})$  measured at CLEO, has been confirmed and made much more accurate by the BaBar collaboration, $\Gamma(D^{\ast +})=(83.5\pm 1.7\pm 1.2)$~keV~\cite{babar}.  

Prior to the CLEO result, the QCD based theoretical estimates of $\gpi$, were made by means of various  QCD sum rules (QSR). The results of different QSR techniques were converging to a small value of $\gpi$~\cite{QCDSR-before,Khodjamirian:1999hb,Colangelo:1994es}, the size of which was challenged in ref.~\cite{with-alain} where it was argued that it should be by about a factor of two larger in order to avoid a paradox of non-saturation  of the Adler-Weisberger sum rule, which is reasonably well saturated by the first couple of states in the case of baryons and light mesons.  As far as the lattice QCD computations  of this coupling are concerned, there have been quite a few results reported in the static heavy quark limit ($m_c\to \infty$)~\cite{deDivitiis:1998kj, static}, while only a few calculations have been reported for the case of  propagating charm quark~\cite{Abada:2002xe,haas,erkol}.~\footnote{While $D^\ast \to D\pi$ decay is allowed, $B^\ast \to B\pi$ decay is kinematically forbidden.}  The first such a computation was performed in quenched approximation in ref.~\cite{Abada:2002xe}, where the strategy to compute $\gpi$ has been discussed in detail. It is based on the use of  the LSZ reduction formula and the axial Ward identity, so that $\gpi$ can be extracted from the computation of the matrix element of the light quark axial current, sandwiched by the $D^\ast$ and $D$ states. The resulting $\gpi$, obtained from the  quenched lattice QCD study ($\nf=0$), was large as anticipated in~\cite{with-alain} and compatible with the CLEO measurement of  $\Gamma(D^{\ast +})$. That study has then been extended to the unquenched case, i.e. by including the effect of $\nf=2$ dynamical quark flavors in the QCD vacuum fluctuations~\cite{haas}. Despite the fact that the ${\cal O}(a)$-improved Wilson quark action was used, the results were obtained for relatively large pion masses and  with a rather poor statistics at two lattice spacings so that the continuum extrapolation was not feasible. In this paper we provide the first result of the unquenched computation of $\gpi$ by using twisted mass QCD on the lattice~\cite{sint,fr}, with $\nf=2$ dynamical light quarks, and at four different lattice spacings. Moreover, our computation is made by using considerably lower pion masses than in ref.~\cite{haas}. Thanks to the high statistics of the ensembles of gauge field configurations that we had at our disposal we were able to make the continuum extrapolation and obtain the physically relevant value of $\gpi$ in the continuum limit.  As we shall see, with respect to the previous lattice estimates, our $\gpi$ coupling is somewhat smaller, but still much larger than the predictions based on using the QSR techniques. While this paper was in writing, an estimate of $\gpi$ at a single value of the lattice spacing with $\nf=2+1$ dynamical  ${\cal O}(a)$-improved Wilson quarks was made in ref.~\cite{erkol} which agrees  very well with our results in the continuum.

The remainder of this paper is organized as follows: we first, in Sec.~\ref{sec:Definitions} define the quantities to compute and specify the correlation functions that are needed for their extraction; 
we then in Sec.~\ref{sec:Lattices} illustrate the quality of the signals obtained from the correlation functions on our lattices and present our results obtained for each of the lattice set-ups; in Sec.~\ref{sec:Results}  we discuss the chiral extrapolations and compare our result with the other results existing in the literature; finally we conclude in Sec.~\ref{sec:Conclusion}.

\section{Definitions\label{sec:Definitions}}
\setcounter{equation}{0}
The coupling $\gpi$ describes the emission of the soft  $P$-wave pion off the $D^\ast$ meson, i.e. it is defined via the following matrix element
\bea
\langle D(k) \pi(q) \vert D^\ast(p,\lambda)\rangle = (e_\lambda \cdot q)\ \gpi \,,
\eea
where $q=p-k$ is the pion momentum and $\lambda$ labels the polarization state of the vector meson.  The decay rate of this process is given by 
\bea\label{eq:width}
\Gamma(D^{\ast}\to D\pi) = {C \over 24 \pi m_{D^{\ast }}^2}\  g_{D^\ast D\pi}^2 \vert \vec k_\pi\vert^3\,,
\eea
where $C=1$ if the outgoing pion is charged, and $C=1/2$ if it is neutral, and  
\bea
 \vert \vec k_\pi\vert ={\lambda^{1/2}(m_{D^\ast},m_D,m_\pi) \over 2 m_{D^\ast} }\,,
\eea
where $\lambda(a,b,c)=[a^2 - (b-c)^2][a^2 - (b+c)^2]$, which numerically gives  $\vert \vec k_{\pi^-}\vert = 39.4\ \mev$ and  $\vert \vec k_{\pi^0}\vert = 38.3\ \mev$, i.e. the pion remains very soft. Due to $T$-symmetry and Lorentz invariance, the same expression applies to the pion absorption, i.e. to $D\pi \to D^\ast$ process.

To compute $\gpi$ we follow the procedure proposed in refs.~\cite{deDivitiis:1998kj,Abada:2002xe}, and compute the matrix element of the axial current, $A_\mu = \bar u \gamma_\mu\gamma_5 d$,
\begin{equation}\label{ffsA}
\begin{split}
\langle &D(k)\vert A^\mu \vert D^\ast (p,\lambda)\rangle = 2m_{D^\ast} A_0(q^2)\frac{\epsilon_\lambda\cdot q}{q^2}q^\mu\\
&+(m_D+m_{D^\ast})A_1(q^2)\left(\epsilon_\lambda^\mu-\frac{\epsilon_\lambda\cdot q}{q^2}q^\mu\right)\\
&+A_2(q^2)\frac{\epsilon_\lambda\cdot q}{m_D+m_{D^\ast}}\left( p^\mu+k^\mu-\frac{m_{D^\ast}^2-m_{D}^2}{q^2}q^\mu\right) \,,
\end{split}
\end{equation}
where $A_{0,1,2}(q^2)$ are three independent Lorentz invariant form factors. For $q^2$ close to $m_\pi^2$ the reduction formula leads to
\begin{equation}\label{lsz}
\begin{split}{f_\pi  m_\pi^2\over m_\pi^2 - q^2} \langle D(k) \pi(q) \vert D^\ast (p,\lambda)\rangle = \langle D(k)\vert \partial_\mu A^\mu \vert D^\ast(p,\lambda)\rangle\, ,
\end{split}
\end{equation}
which for $q^2=0$ gives, 
\bea\label{eq:01}
g_{D^\ast D\pi} = {2 m_{D^\ast}\over f_\pi} A_0(0)\,.
\eea
Since the form factor $A_0(q^2)$ is dominated by pion ($J^P=0^-$) in the $t$-channel, it strongly varies with $q^2$, as well as with the light quark mass when considering the problem for the practical lattice QCD calculation. 
However, since no massless state can couple to the axial current, one has
\bea
A_0(0)={m_{D^\ast}+m_D\over 2m_{D^\ast} } A_1(0)-{m_{D^\ast}-m_D \over 2m_{D^\ast} }A_2(0) \,,
\eea
so that eq.~(\ref{eq:01}) becomes
\begin{equation}\label{defA}
\begin{split} 
g_{D^\ast D\pi}= \frac{m_{D^\ast}+m_D}{f_\pi} A_1(0)\left[1+{m_{D^\ast}-m_D\over m_{D^\ast}+m_D}{A_2(0)\over A_1(0)}\right]  .
\end{split}
\end{equation}
The last expression is suitable for numerical computations on the lattice. It reduces to the calculation of the following correlation functions
\bea\label{eq:00}
C_{\mu\nu}(\vec q;t)= \sum_{\vec x,\vec y } \langle V_\mu(\vec  0,0) A_\nu (\vec x, t)   P_5^\dag (\vec y, t_S) e^{-i\vec q (\vec x-\vec y)}\rangle\,, 
\eea
where the interpolating source operators $V_\mu$ and $P_5$ are placed far from each other at $t=0$ and $t=t_S$, and create the $D^\ast$- and $D$-meson states, respectively, at  $t$ such that $0\ll t\ll t_S$.   
The simplest choice are the local interpolating operators, $V_\mu=\bar c\gamma_\mu u$ and $P_5=\bar c\gamma_5 d$. Note also that $\vec q$ should be chosen in such a way that 
the form factors are computed at $q^2=0$, as needed in eq.~(\ref{defA}). If the vector meson is kept at rest, that means that the desired three-momentum should be tuned to
\bea\label{eq:02}
\vert \vec q\vert ={ m_{D^\ast}^2 - m_{D}^2\over 2 m_{D^\ast} }\,,
\eea
which is too small a value to be accommodated on the periodic lattices explored in the current numerical simulations. 
To remedy this difficulty one of the quark propagators, $S_q(x,0;U) \equiv \langle q(x)\bar q(0)\rangle$, can be computed 
  in the rephased gauge field configuration,  
\bea\label{twisted}
U_\mu(x) \to U^\theta_\mu(x) = e^{i \theta_\mu/L}U_\mu(x)\ ,
\eea
where $ \theta_\mu = (0,\vec \theta)$, and $L$ is the size of the spatial side of the cubic box. The propagator  
\bea\label{eq:tbc}
S_q^{ \vec \theta} (x,0;U)  = e^{i \vec \theta\cdot \vec x/L} \ S_q(x,0;U^\theta) \,,
\eea
is equivalent to imposing the twisted boundary conditions on one of the valence quarks in the considered correlation function~\cite{nazario,chris-giovanni} (see also ref.~\cite{diego}).  
Importantly, however,  $\theta_0$ in $\vec \theta = (\theta_0, \theta_0,\theta_0)$ can be tuned to any value, and therefore also to 
\bea\label{eq:theta0}
\theta_0 = {L\over \sqrt{3}} \vert \vec q\vert = {L\over \sqrt{3}}  { m_{D^\ast}^2 - m_{D}^2\over 2 m_{D^\ast} } \,,
\eea
which ensures the extraction of the form factors at $q^2=0$. For such a tuned $\theta_0$, and by using the local interpolating field operators, the correlation function~(\ref{eq:00}) reads,
\begin{equation}\label{eq:three-tw}
\begin{split} 
&C_{\mu\nu}(\vec q;t) = \\
& \langle \sum_{\vec x,\vec y}{\rm Tr}\left[  \gamma_\mu S_c(0,y) \gamma_5 S^{ \vec \theta}_d (y,x;U)\gamma_\nu \gamma_5 S_u(x,0;U)  \right]\rangle \,, 
\end{split}
\end{equation}
which is very close to what we actually computed on the lattice.~\footnote{As we shall see in Sec.~\ref{sec:Lattices} the difference with respect to eq.~(\ref{eq:three-tw}) is that in practice we use the extended (smeared) source operators.} To relate the above correlation function to the relevant form factors we note that 
\begin{equation} \label{eq:G10}
\begin{split} 
\widetilde C_{i j}(\vec q;t) = {1\over 3}  \sum_{i=1}^3 C_{i i}(\vec q;t) - {1\over 6}  \sum_{i,j=1}^3 \biggl.C_{i j}(\vec q;t)\biggr|_{i\neq j},
\end{split}
\end{equation}
at large time separations among operators, behaves as
\begin{equation} 
\begin{split} \label{eq:G11}
\widetilde C_{i j}(\vec q;t) \to {{\cal Z}_{D^\ast}  \over 2 m_{D^\ast} } e^{- m_{D^\ast}t} &\times (m_D + m_{D^\ast}) A_1(0) \\ &\times {{\cal Z}_{D}  \over 2 E_{D} } e^{- E_{D} (t_S-t)} ,
\end{split}
\end{equation}
where $\langle 0\vert P_5\vert D\rangle = {\cal Z}_{D}$, and  $\langle 0\vert V_i\vert D^\ast \rangle = e_i^\lambda {\cal Z}_{D^\ast}$. In this way we get the first term in eq.~(\ref{defA}). To reach the second term within the brackets 
in (\ref{defA}) we proceed along the lines explained in ref.~\cite{haas} and compute
\begin{equation} 
\begin{split} 
\widetilde C_{i0}(\vec q;t) =&- {1\over 3}  \sum_{i=1}^3  C_{i0}(\vec q;t)\\
&+\biggl. {1\over 6}  \sum_{i,j=1}^3 \displaystyle{\frac{m_{D^\ast} - E_D}{ q_i}} C_{i j}(\vec q;t) \biggr|_{i\neq j}
\end{split}
\end{equation}
so that at large separations among the operators, one gets
\bea\label{eq:ct}
{ \widetilde   C_{i0}(\vec q;t)  \over  \widetilde C_{i j}(\vec q;t)  } \to  {2 q_i m_{D^\ast}\over 
 (m_{D^\ast} + m_D)^2 } \ {A_2(0)\over A_1(0)}\,,
\eea
where $q_i=\theta_0/L$. Hadron masses, $m_{D^{(\ast)}}$, and couplings, ${\cal Z}_{D^{(\ast)}}$, are extracted from the study of the large time behavior of the two-point  correlation functions, namely
\begin{equation} \label{eq:2pts}
\begin{split} 
 \langle {\displaystyle \sum_{\vec x} }  V_{i}&(\vec x; t)  V^\dagger_{i}(0; 0) \rangle  \xrightarrow[]{\displaystyle{ t\gg 0}}  \\
 &\;  \left| {\cal Z}_{D^\ast} \right|^2 \frac{\cosh[  m_{D^\ast} (T/2-t)]}{ m_{D^\ast_q} } e^{- m_{D^\ast} T/2}\,,\\
 \langle {\displaystyle \sum_{\vec x} }  P_5&(\vec x; t)  P_5^\dagger (0; 0) \rangle  \xrightarrow[]{\displaystyle{ t\gg 0}}  \\
 &\;  \left| {\cal Z}_{D} \right|^2 \frac{\cosh[  m_{D} (T/2-t)]}{ m_{D^\ast_q} } e^{- m_{D} T/2}\,,
\end{split}
\end{equation}
where $T$ stands for the size of the temporal extension of the lattice. Note that in eq.~(\ref{eq:2pts}) we used the symmetry of the correlation functions with respect to $t\leftrightarrow T-t$.

\section{Lattices used in this work\label{sec:Lattices}}

Results presented in this work are obtained by using the gauge field configurations generated by the European Twisted Mass Collaboration (ETMC)~\cite{Boucaud:2008xu}, by simulating the twisted mass QCD on the lattice at the maximal twist~\cite{fr}. $\nf=2$ dynamical light quarks have been included in the simulations. 
In tab.~\ref{tab:01} we collect the main information concerning the lattices used in this work. At each lattice spacing, simulations have been made with several values of the sea quark mass $\mu_{\rm sea}$,  covering the range of the associated pion masses $m_\pi \in (280, 500)$~MeV~\cite{Blossier:2010cr}. Concerning the charm quark mass, its value has been discussed in great detail in ref.~\cite{Blossier:2010cr} 
and it is fixed by requiring the agreement between the charmed hadron masses ($D$, $D_s$ and $\eta_c$ mesons) computed on the lattice with their  experimentally established values. 

Our first task was to compute the two-point functions~(\ref{eq:2pts}) and extract the accurate values of the masses of $D_q$ and $D_q^\ast$ mesons 
which are necessary for determination of  $\theta_0$ in eq.~(\ref{eq:theta0}).~\footnote{Index ``$q$" in $D_q^{(\ast)}$ labels the light valence quark, which in our computations is always kept degenerate in mass with the sea quark.} In the computation of the  two-point functions~(\ref{eq:2pts})  we used source operators composed of local quark fields
$\psi$ or smeared ones $\psi_{n_g}$.  We choose the gaussian smearing defined via:
\begin{equation}
\psi_{n_g}=\left(\frac{1+\kappa H}{1+6\kappa}\right)^{n_g}\psi\,,
\end{equation}
where $H$ is the smearing operator defined as~\cite{Gusken:1989ad}
\begin{equation}
H_{i,j}=\sum_{\mu=1}^3\left(U^{n_a}_{i;\mu}\delta_{i+\mu,j}+U^{n_a\dagger}_{i-\mu;\mu}\delta_{i-\mu,j}\right)\,,
\end{equation}
with $U^{n_a}_{i,\mu}$ being the $n_a$ times APE smeared link~\cite{Albanese:1987ds}, defined in terms of $(n_a-1)$ times smeared link $U^{(n_a-1)}_{i,\mu}$ and its surrounding staples $V^{(n_a-1)}_{i,\mu}$, 
\begin{equation}
U^{n_a}_{i,\mu}={\rm Proj_{SU(3)}}\left[(1-\alpha)U^{(n_a-1)}_{i,\mu}+\frac{\alpha}{6} V^{(n_a-1)}_{i,\mu}\right ]\,.
\end{equation}
For the smearing parameters we choose:
\begin{equation}
\kappa=4,\,n_g=30,\,\alpha=0.5,\,n_a=20\,,
\end{equation}
which we kept fixed for all of our lattices. In our calculation the quark propagators are computed by using stochastic sources,
and the correlation function are obtained taking advantage of the so called one-end trick~\cite{Boucaud:2008xu}.
Furthermore, to improve the statistical quality of our results, we use up to eight different spatial sources. 
\begin{table*}[h!!]
\begin{ruledtabular}
\begin{tabular}{|c|cccccc|}   
{\phantom{\huge{l}}}\raisebox{-.2cm}{\phantom{\Huge{j}}}
$ \beta$& 3.8 &  3.9  &  3.9 & 4.05 & 4.2  & 4.2    \\ 
{\phantom{\huge{l}}}\raisebox{-.2cm}{\phantom{\Huge{j}}}
$ L^3 \times T $&  $24^3 \times 48$ & $24^3 \times 48$  & $32^3 \times 64$ & $32^3 \times 64$& $32^3 \times 64$  & $48^3 \times 96$  \\ 
{\phantom{\huge{l}}}\raisebox{-.2cm}{\phantom{\Huge{j}}}
$ \#\ {\rm meas.}$& 240 &  240 & 144 & 144 & 144 & 96  \\ \hline 
{\phantom{\huge{l}}}\raisebox{-.2cm}{\phantom{\Huge{j}}}
$\mu_{\rm sea 1}$& 0.0080 & 0.0040 & 0.0030 & 0.0030 & 0.0065 &  0.0020   \\ 
{\phantom{\huge{l}}}\raisebox{-.2cm}{\phantom{\Huge{j}}}
$\mu_{\rm sea 2}$& 0.0110 & 0.0064 & 0.0040 & 0.0060 &   &     \\ 
{\phantom{\huge{l}}}\raisebox{-.2cm}{\phantom{\Huge{j}}}
$\mu_{\rm sea 3}$&  & 0.0085 &  & 0.0080 &   &     \\ 
{\phantom{\huge{l}}}\raisebox{-.2cm}{\phantom{\Huge{j}}}
$\mu_{\rm sea 4}$&  & 0.0100 &  &   &   &     \\   \hline 
{\phantom{\huge{l}}}\raisebox{-.2cm}{\phantom{\Huge{j}}}
$a \ {\rm [fm]}$&   0.098(3) & 0.085(3) & 0.085(3) & 0.067(2) & 0.054(1) & 0.054(1)      \\ 
{\phantom{\huge{l}}}\raisebox{-.2cm}{\phantom{\Huge{j}}}
$Z_A (g_0^2)$~\cite{Blossier:2010cr,Constantinou:2010gr}& 0.746(11) & 0.746(6) & 0.746(6)  & 0.772(6) & 0.780(6)& 0.780(6) \\ 
{\phantom{\huge{l}}}\raisebox{-.2cm}{\phantom{\Huge{j}}}
$\mu_{c}$& 0.2331(82)  &0.2150(75)  &0.2150(75)   & 0.1849(65) & 0.1566(55) & 0.1566(55)  \\ 
\end{tabular}
{\caption{\footnotesize  \label{tab:01} Lattice ensembles used in this work with the indicated number of gauge field configurations. Lattice spacing is fixed by using the Sommer parameter $r_0/a$~\cite{R0}, with $r_0= 0.440(12)$~fm fixed by matching $f_\pi$ obtained on the lattice with its physical value  (c.f. ref.~\cite{Blossier:2010cr}). Quark masses are given in lattice units.}}
\end{ruledtabular}
\end{table*}
The full list of results is presented in tab.~\ref{tab:02} where we give our values for $m_{D_q}$ and $m_{D^\ast_q}$ that are totally consistent with our previous findings presented 
in ref.~\cite{Becirevic:2012ti} and obtained by using the single source propagators. Once we have the meson masses we fix the value of $\theta_0$ according to  eq.~(\ref{eq:theta0}), and then compute the light quark propagator with twisted 
boundary conditions~(\ref{eq:tbc}), which is needed to evaluate the three-point functions~(\ref{eq:three-tw}).

\begin{figure}[t!]
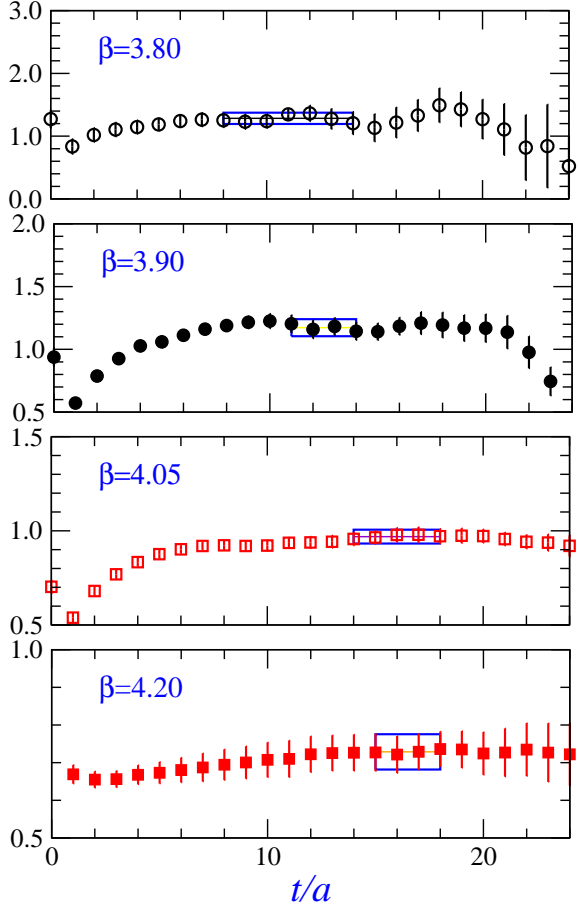

\epsfig{file=plateaux-380.eps, height=2.8cm}\\
\epsfig{file=plateaux-390.eps, height=2.8cm}\\
\epsfig{file=plateaux-405.eps, height=2.8cm}\\
\epsfig{file=plateaux-420.eps, height=3.53cm}\\
\caption{\label{fig:1}{\sl 
Ratio $R_1(t)$, defined in eq.(\ref{eq:R1}) [see also eqs.~(\ref{eq:G10},\ref{eq:G11})], as a function of time, with two sources of $D_q^\ast$ and $D_q$ mesons fixed. Illustration is provided at each of the four lattice spacings considered and with the light quark mass  $\mu_q=0.008$, $0.004$, $0.008$, and  $0.002$, for $\beta =3.80$, $3.90$, $4.05$, and $4.20$, respectively. Also indicated is the plateau region at which a fit to a constant has been made, $R_1(t) \to (m_{D_q}+ m_{D_q^\ast}) A_1(0)$.}} 
\end{figure}

Instead of the coupling $\gpi$, it is often convenient to define $g_c$ which is related to $\gpi$ via
\bea\label{eq:gc}
\gpi = {2\sqrt{m_D m_{D_\ast}}\over f_\pi} g_c\,,
\eea
with the convention that corresponds to $f_\pi = 130.41(3)(20)$MeV~\cite{PDG}. On the basis of heavy quark expansion the coupling $g_c$ is expected to scale as a constant up to corrections proportional to the powers of the inverse heavy quark (meson) mass. In our case, we work with the propagating heavy charm quark, which is why we added an index ``$c$" to avoid ambiguities when comparing $g_c$ computed here with the values obtained in the static limit~\cite{static}. After combining eq.~(\ref{defA}) with (\ref{eq:gc}) we have
\begin{equation}\label{gc:final}
\begin{split}
g_c&= \frac{m_{D^\ast}+m_D}{2\sqrt{m_D m_{D_\ast}}} A_1(0)\left[1+{m_{D^\ast}-m_D\over m_{D^\ast}+m_D}{A_2(0)\over A_1(0)}\right]  \\
&= G(0) \left[ 1 + G^\prime (0)\right]\,,
\end{split}
\end{equation}
where $G^{(\prime )}(0)$ are introduced for shortness. The values of the momenta tuned in the twisted boundary conditions~(\ref{eq:theta0}) are very small and by computing $G^\prime(0)$ by using eq.~(\ref{eq:ct}) its value is consistent with zero for all our lattices. The leading term, $G(0)$, instead is extracted  accurately from the correlation functions~(\ref{eq:G10}). In fig.~\ref{fig:1} we illustrate the plateaus obtained at all of our four lattice spacings and for one value of the sea quark mass. Illustrated is the ratio
\bea
\label{eq:R1}
R_1(t) &=&{ 2 E_{D} 2 m_{D^\ast} \widetilde C_{i j}(\vec q;t) \over {\cal Z}_{D}  {\cal Z}_{D^\ast} } e^{ m_{D^\ast}t + E_{D} (t_S-t)} \cr
 &&\xrightarrow[]{{t_S\gg t\gg 0}}  (m_{D_q}+ m_{D_q^\ast}) A_1(0)\,.
\eea
Corresponding values of $g_c$ are given in tab.~\ref{tab:02} for all of our lattice set-ups. Fits are made on the plateaus corresponding to 
$t_\beta \in [8,14]_{3.9}$, $[11,14]_{4.05}$, $[15,18]_{4.2}$, in an obvious notation.  Those plateaus are chosen to be common to all sea quark mass values considered at a given lattice spacing.  

We reiterate that in our computations the sea and the valence light quarks are always kept degenerate in mass, and the corresponding $m_\pi L \gtrsim 4$ is kept large to make the finite volume effects small. From the simulations made at $\beta=3.9$ with $\mu_q=0.004$ we were able to verify that the finite volume effects for $g_c$ are indeed small by comparing its values  obtained on the $24^3\times 48$ and on the $32^3\times 64$ lattices.  Within the given accuracy they are completely consistent and the finite volume effects are neglected in what follows.
\begin{table*}[h!!]
\begin{ruledtabular}
\begin{tabular}{|ccc|c|c|c|c|} 
{\phantom{\huge{l}}} \raisebox{-.2cm} {\phantom{\huge{j}}}
$\beta$ & $L$  & \quad $\mu_q$   & $ m_q^\msbar (2\,\gev)$   & $ m_{D_q}$  & $ m_{D^\ast_q} $       & $g_c$                \\ \hline\hline
{\phantom{\huge{l}}} \raisebox{-.2cm} {\phantom{\huge{j}}}
3.80 & 24 & \qquad$\mu_{\rm sea 1}$ &  0.0398(11)& 1.73(5) & 1.95(6) & 0.536(36) \\  
{\phantom{\huge{l}}} \raisebox{-.2cm} {\phantom{\huge{j}}}          
     &    & \qquad$\mu_{\rm sea 2}$ & 0.0547(15) & 1.75(5) & 1.98(6) & 0.524(26) \\ \hline
{\phantom{\huge{l}}} \raisebox{-.2cm} {\phantom{\huge{j}}}          
3.90 & 24 & \qquad$\mu_{\rm sea 1}$ & 0.0216(5)  & 1.75(5) & 1.96(5) & 0.559(33) \\ 
{\phantom{\huge{l}}} \raisebox{-.2cm} {\phantom{\huge{j}}}          
  &   &\qquad$\mu_{\rm sea 2}$      & 0.0345(8)  & 1.77(5) & 1.98(5) & 0.564(24) \\  
{\phantom{\huge{l}}} \raisebox{-.2cm} {\phantom{\huge{j}}}          
   &  &\qquad$\mu_{\rm sea 3}$      & 0.0458(11) & 1.78(5) & 1.99(5) & 0.589(18) \\
{\phantom{\huge{l}}} \raisebox{-.2cm} {\phantom{\huge{j}}}          
   &  &\qquad$\mu_{\rm sea 4}$      & 0.0539(13) & 1.78(5) & 1.99(5) & 0.588(13) \\ \hline
{\phantom{\huge{l}}} \raisebox{-.2cm} {\phantom{\huge{j}}}          
3.90  & 32 & \qquad$\mu_{\rm sea 1}$& 0.0162(4)  & 1.74(5) & 1.94(5) & 0.509(24) \\ 
{\phantom{\huge{l}}} \raisebox{-.2cm} {\phantom{\huge{j}}}          
    &  &\qquad $\mu_{\rm sea 2}$    & 0.0216(5)  & 1.75(5) & 1.95(5) & 0.543(37) \\ \hline
{\phantom{\huge{l}}} \raisebox{-.2cm} {\phantom{\huge{j}}}          
4.05  & 32 &\qquad $\mu_{\rm sea 1}$ & 0.0249(7) & 1.81(4) & 2.01(4) & 0.528(29) \\ 
{\phantom{\huge{l}}} \raisebox{-.2cm} {\phantom{\huge{j}}}          
  &   &\qquad $\mu_{\rm sea 2}$     & 0.0374(10) & 1.83(4) & 2.03(5) & 0.585(22) \\ 
{\phantom{\huge{l}}} \raisebox{-.2cm} {\phantom{\huge{j}}}          
    & &\qquad $\mu_{\rm sea 3}$     & 0.0499(13) & 1.84(4) & 2.04(5) & 0.575(22) \\ \hline
{\phantom{\huge{l}}} \raisebox{-.2cm} {\phantom{\huge{j}}}          
4.20  & 32 &\qquad $\mu_{\rm sea 1}$ & 0.049(2)  & 1.91(4) & 2.11(5) & 0.620(18) \\ \hline
{\phantom{\huge{l}}} \raisebox{-.2cm} {\phantom{\huge{j}}}          
4.20  & 48 &\qquad $\mu_{\rm sea 1}$ & 0.0150(7) & 1.87(4) & 2.06(4) & 0.540(35) \\ 
\end{tabular}

{\caption{  \label{tab:02} \sl
$D_q$ and $D^\ast_q$ meson masses obtained with the charm quark mass fixed to its physical value and the light quark mass equal to the mass of the sea quark, specified in tab.~\ref{tab:01}. 
and  for various light quark masses. The values for $g_c$ are obtained as described in the text. All masses are given in physical units [GeV], and $g_c$ is dimensionless. }}
\end{ruledtabular}
\end{table*}

Finally we should also mention that the renormalization factors of the axial current, $Z_A(g_0^2)$, have been properly accounted for in our computation. Their values have been computed non-perturbatively in refs.~\cite{Blossier:2010cr,Constantinou:2010gr} and, for convenience,  they are also given in tab.~\ref{tab:01} of the present paper.

\section{Physical results and phenomenology\label{sec:Results}}
\subsection{Getting to the physical value for $g_c$}
The values of the coupling $g_c$, obtained with unphysical light quark masses from our lattices, should now be related to the physically relevant coupling $g_c$.  
To do so we will make the continuum and the chiral extrapolations simultaneously by employing three different chiral extrapolation formulas. Concerning the continuum extrapolation we will rely on the simple linear extrapolation in $a^2$, which turns out to be adequate for describing our data. Concerning the chiral extrapolation we will use the simple linear extrapolation in the light quark mass (or, equivalently, in the pion mass squared), and the two formulas that include the chiral logarithmic corrections. In other words we will use:
\begin{itemize}
\item Linear extrapolation: 
\bea \label{eq:extrap-1}
 g_c=g_0 \left( 1 + \alpha m_\pi^2 + \beta {a^2\over a_{3.9}^2} \right)\,, 
\eea
where on the left hand side are the values we obtain on our lattices for a given light quark mass (pion mass), in which the valence and sea quark mass are the same. 
On the right hand side, $g_0$ stands for the same coupling obtained in the chiral limit, and $\alpha$ is the slope parameter. 
Notice also that we divide $a^2$ by the value of the lattice spacing obtained at $\beta=3.9$, so that the $\beta$ parameter is a measure of the deviation of our continuum result form the values obtained at $a=0.085(3)$~fm.
\item Chiral Perturbation Theory  (ChPT-I):
\begin{equation} \label{eq:extrap-2}
\begin{split}
 g_c=g_0 \left( 1- {4 g_0^2\over (4\pi f)^2} m_\pi^2 \log m_\pi^2 + \alpha m_\pi^2 + \beta {a^2\over a_{3.9}^2} \right)\,, 
\end{split}
\end{equation}
is the formula that has been derived many times in the heavy meson chiral perturbation theory (HMChPT)~\cite{HMChPT}, an effective theory constructed from the combined heavy quark symmetry with the spontaneous chiral symmetry breaking pattern in order to describe the mesons consisting of heavy and light quarks~\cite{jernej,Detmold:2011rb}. The above formula~(\ref{eq:extrap-2}) is obtained at the next-to-leading order in chiral expansion and to the leading order in the heavy quark expansion (i.e. in the static limit of QCD).  Although we are working with the charm heavy quark, for which the above formula is not fully adequate, we will use it as a way to guide our extrapolation to the physical coupling $g_c$. The constant $f$ in the above formula stands for the pion decay constant in the chiral limit, which we take to be $f=120$~MeV. 
\begin{figure}[h!]
\epsfig{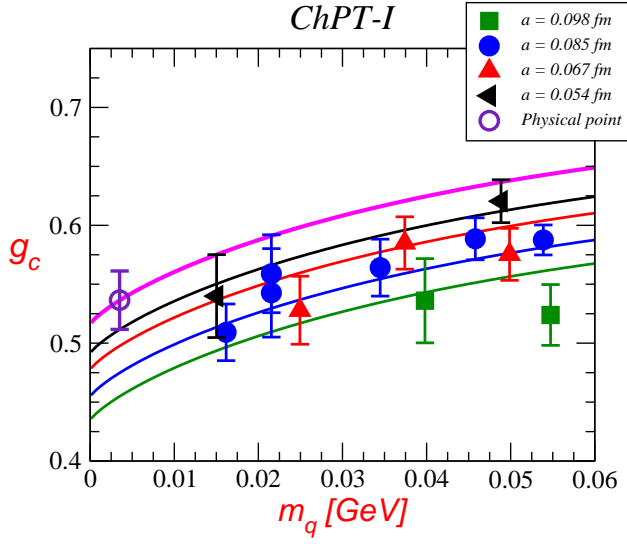}\\
\caption{\label{fig:2}{\sl 
Chiral and continuum extrapolation of our lattice QCD results for $g_c$ given in tab.~\ref{tab:02}, by using eq.~(\ref{eq:extrap-2}).}} 
\end{figure}
In fig.~\ref{fig:2} we show the result of extrapolation guided by eq.~(\ref{eq:extrap-2}). For the light quark mass we use the definition $m_q^\msbar (2 \ \gev)$ in the same way it was defined (renormalized) in ref.~\cite{Blossier:2010cr}. As usual, $m_\pi^2= 2 B_0 m_q$, and $B_0^\msbar(2\ \gev)=2.71(15)$~GeV.
\item Chiral Perturbation Theory (ChPT-II):
\begin{equation} \label{eq:extrap-3}
\begin{split}
 g_c=g_0 \left( 1- {2 (1+2 g_0^2)\over (4\pi f)^2} m_\pi^2 \log m_\pi^2 + \alpha m_\pi^2 + \beta {a^2\over a_{3.9}^2} \right)\,, 
\end{split}
\end{equation}
was derived in ref.~\cite{Detmold:2011rb} where it was noted that the expression~(\ref{eq:extrap-2}) was obtained by using the pion field directly, and not the axial current. In principle, our formula~(\ref{eq:01}) is related to the pion field directly. However, as we mentioned in Sec.~\ref{sec:Lattices}, our result is completely dominated by the axial form factor $A_1(0)$, and therefore the formula~(\ref{eq:extrap-3}), which differs from~(\ref{eq:extrap-2}) by the vertex tadpole diagram, should be used in  chiral extrapolation too.  Notice that in both  eqs.~(\ref{eq:extrap-2},\ref{eq:extrap-3})  the $\mu$-dependence of the log term and of the counter-term coefficient $\alpha$ are not written explicitly as they cancel in the sum. Notice also that the expressions in eqs.~(\ref{eq:extrap-2},\ref{eq:extrap-3}) refer to a theory with $\nf=2$ flavors, based on the spontaneous chiral symmetry breaking $SU(2)_L\otimes SU(2)_R \to SU(2)_V$.
\begin{figure}[h!]
\epsfig{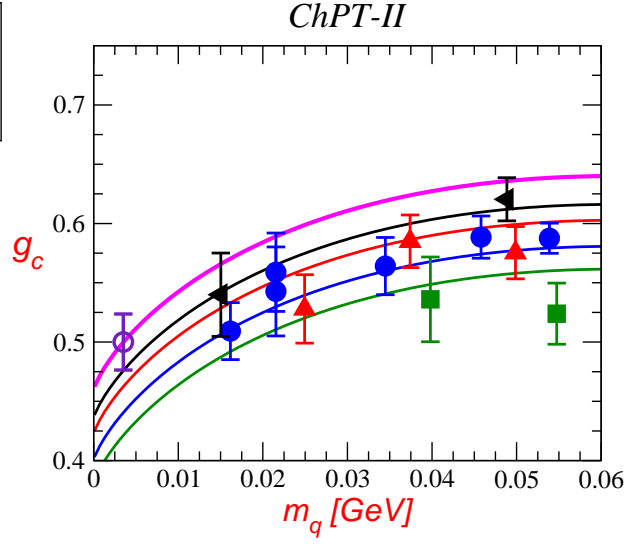}\\
\caption{\label{fig:3}{\sl 
Same as in fig.~\ref{fig:2} but using eq.~(\ref{eq:extrap-3}).}} 
\end{figure}
\end{itemize}
We stress again that the above expressions are derived in the static heavy quark limit. It is not clear to what extent the chiral logarithmic corrections are relevant in describing the light quark dependence of the heavy-light hadronic quantities, and it is even less clear whether or not the above expressions could be applied to the case of mesons with (not so heavy) charm quark.  In other words, none of the above formulas is exactly applicable to our situation and we will use all of them. The spread of results associated with each of the formulas~(\ref{eq:extrap-1},\ref{eq:extrap-2},\ref{eq:extrap-3})  could then be used to estimate the systematic uncertainty due to the chiral extrapolation. Note however that the results of refs.~\cite{1surM} suggest that the chiral logarithmic behavior of the $g$-coupling remains unaltered after including the leading power corrections.

That  error will leave enough room for the future improvement  of $g_c$ (i.e. of $\gpi$) computed on the lattice.

In more detail, from the fit of our data to the above formulas we obtain:
\begin{align}
{\underline{\rm Linear}}&\cr
&\hfill \cr
 g_0 =0.56(3),&\alpha= 0.6(2)\ \gev^{-2}, \;  \beta=-0.12(4),\cr
& \Rightarrow \,   g_c^{\rm phys.} = 0.562(28)\,, 
\end{align}
\begin{align}
{\underline{\rm ChPT-I}}&\cr
&\hfill \cr
  g_0 =0.52(2),& \alpha= 0.3(2)\ \gev^{-2}, \; \beta=-0.12(4),\cr
& \Rightarrow \,  g_c^{\rm phys.} = 0.536(25)\,, 
\end{align}
\begin{align}
{\underline{\rm ChPT-II}}&\cr 
&\hfill \cr
g_0 =0.46(2),&\alpha= -0.16(25)\ \gev^{-2}, \;  \beta=-0.13(5) ,\cr
&\quad  \Rightarrow \,   g_c^{\rm phys.} = 0.500(24)\,.
\end{align}
As our final result we quote
\bea\label{eq:final}
g_c=0.54(3)(^{+2}_{-4})\,,
\eea
where the errors refer to the last digit. The first error is statistical and is to be understood as gaussian, while the second one is uniform and accounts for the spread of values obtained by using various fit procedures to reach the physical limit. 

\subsection{Phenomenology}
The above coupling~(\ref{eq:final}) can now be cast into the form in which it is usually discussed in phenomenology by using eq.~(\ref{eq:gc}). We obtain,
\bea
\gpi=g_{D^{\ast +} \overline D^0\pi^+} = 15.9(7)(^{+2}_{-4})\,,
\eea
which then leads us to our estimate for the decay width~(\ref{eq:width}), namely,
\bea
\Gamma(D^{\ast +}\to \overline D^0\pi^+) = \left(51\pm 5^{+5}_{-7}\right)~{\rm keV}\,.
\eea
The width of the charged $D^\ast$-meson can be calculated as 
\begin{equation}\label{eq:d+}
\begin{split}
\Gamma(D^{\ast +})=&  \Gamma(D^{\ast +} \to \overline D^0 \pi^+) + \Gamma(D^{\ast +} \to D^+ \pi^0) \cr
&+ \Gamma(D^{\ast +} \to D^+ \gamma)\,,
\end{split}
\end{equation}
as these are the only kinematically available decay channels. The last mode, in particular, is a tiny fraction of the full decay width and is known experimentally,  $B(D^{\ast +}\to D^+\gamma) = 0.016(4)$~\cite{PDG}, so that by using our value for $g_c$ we finally have
\bea
\Gamma(D^{\ast +})  = \left(76 \pm 7^{+8}_{-10}\right)~{\rm keV}\,,
\eea
which agrees favorably with the experimental values, $\Gamma(D^{\ast +})^{\rm CLEO}=96\pm 22$~keV~\cite{cleo}, and $\Gamma(D^{\ast +})^{\rm BaBar}=(83.5\pm 1.7\pm 1.2)$~keV~\cite{babar}.

Furthermore, we can make a prediction of the width of the neutral $D^\ast$ meson,
\bea
\Gamma(D^{\ast 0})&=&  \Gamma(D^{\ast 0} \to D^- \pi^+) + \Gamma(D^{\ast 0} \to D^0 \pi^0) \nn\\
&&+ \Gamma(D^{\ast 0} \to D^0 \gamma)\,,
\eea
if we use the experimental information $B(D^{\ast 0}\to D^0\gamma) = 0.381(29)$~\cite{PDG}. Knowing that $D^{\ast 0} \to D^- \pi^+$ is kinematically forbidden, and assuming the isospin symmetry to relate our $g_c$ to the decay to a neutral pion,  
we obtain
\begin{align}\label{eq:D0}
\Gamma(D^{\ast 0}) [1- B(D^{\ast 0}\to &D^0\gamma)]=  {  m_{D^0}  |\vec k_{\pi^0}^\prime|^3\over 12 \pi\ m_{D^{\ast 0}}\ f_\pi^2 }   g_c^2\nn\\ 
&= \left(53\pm 5^{+6}_{-7}\right)\ \kev\,,
\end{align}
where $|\vec k_{\pi^0}^\prime|=42.3$~MeV, and the prime is used to distinguish this case, in which all mesons are neutral. Our result~(\ref{eq:D0}) is much smaller than the relatively old experimental bound, $\Gamma(D^{\ast 0})< 2.1\ \mev$~\cite{abachi}.

\subsection{Comparison with results available in the literature}

\begin{figure}
\psfrag{AA1}{\color{blue}\large ref. [2]}
\psfrag{AA2}{\color{blue}\large ref. [1]}
\psfrag{BB1}{\color{blue}\sf \large this work}
\psfrag{BB2}{\color{blue}\large ref. [13]}
\psfrag{BB3}{\color{blue}\large ref. [10]}
\psfrag{BB4}{\color{blue}\large ref. [10]}
\psfrag{BB5}{\color{blue}\large ref. [9]}
\psfrag{CC1}{\color{blue}\large ref. [5]}
\psfrag{CC2}{\color{blue}\large ref. [4]}
\epsfig{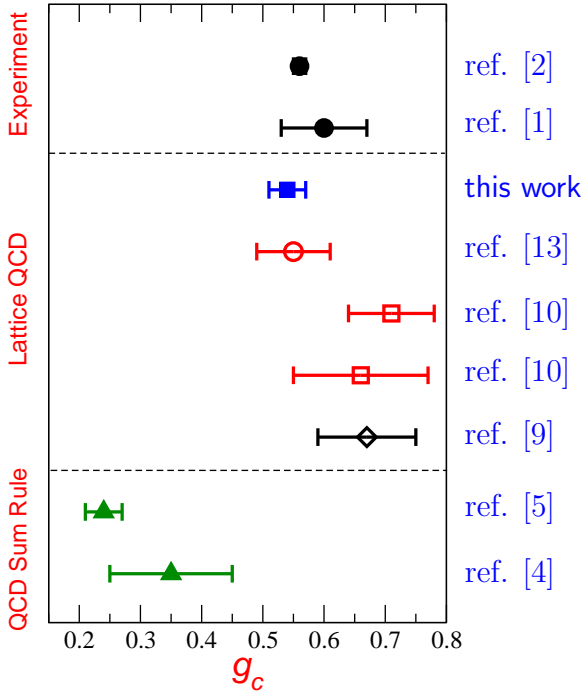}\\
\caption{\label{fig:X}{\sl 
Value of $g_c$ as inferred from experimentally measured $\Gamma(D^{\ast \pm})$, from the QCD simulations on the lattice (full symbol is in the continuum, empty symbols refer to the fixed lattice spacing: diamond for $\nf=0$, square for $\nf=2$ and circle for $\nf=2+1$), and by means of the QCD sum rule technique, after including the radiative corrections. }} 
\end{figure}

To compare our result~(\ref{eq:final}) with the existing experimental ones  we use the expression~(\ref{eq:d+}) and from the  width reported by CLEO~\cite{cleo} we obtain $g_c=0.60(6)$. From the new BaBar result~\cite{babar}, instead, we get  $g_c=0.56(1)$. The lattice QCD results reported so far were obtained by using the ${\cal O}(a)$-improved Wilson quarks at single lattice spacing. The result obtained in the quenched approximation,  $g_c=0.67(8)(^{+4}_{-6})$~\cite{Abada:2002xe}, 
was later confirmed in the unquenched study with $\nf=2$ dynamical light flavors, at nearly the same value of the lattice spacing, $g_c=0.66(11)$~\cite{haas}. At a slightly finer lattice, $g_c=0.71(7)$ was obtained~\cite{haas}. However, all these results  were obtained by working directly with large pion masses. Very recently the result of the lattice QCD study with ${\cal O}(a)$-improved Wilson quarks and with $\nf=2+1$ dynamical light flavors was reported in ref.~\cite{erkol}. Although working at the single lattice spacing the direct computations were made for a broad range of pion masses down to $300$~MeV. The authors fit their result linearly and quadratically to extrapolate to the physical pion mass, without relying on the expressions derived in HMChPT. Their final result, $g_c = 0.55(6)$, agrees well with ours, as well as with the experimental ones. 
Current situation concerning the estimate of $g_c$ is shown in fig.~\ref{fig:X}, where we observe a large difference with respect to the results obtained by using QSR. Ways to remedy this difficulty have been proposed in the literature~\cite{depanage} but a real reason for this failure has not been fully understood. Here we only focused on the computations closely related to QCD. A number of results based on quark model calculations have been reported in the literature too. For an exhaustive list of references see~\cite{with-alain,ElBennich:2010ha}. Extracting this coupling from the residuum of the semileptonic form factor was shown to be very difficult because the relevant $D\to \pi$, $B\to \pi$, or $D\to K$  semileptonic form factor is a strongly varying functions at large $q^2$-region and the extraction of $g_c$ (or $g_b$) significantly depends on the parameterization used to describe the $q^2$-dependence of the form factor~\cite{DescotesGenon:2008hh,Li:2010rh}.

\section{Summary and perspectives\label{sec:Conclusion}}

In this paper we report on the first lattice QCD result for the $g_{D^\ast D\pi}$ coupling obtained in the continuum limit. By using twisted mass QCD on the lattice at four different lattice spacings and by working with $\nf=2$ dynamical light quark flavors, we were able to compute this coupling for several values of the pion mass, covering the range between $280$~MeV and $500$~MeV. Our final value is 
\bea
g_c=0.53(3)(3),\ \Leftrightarrow \ g_{D^{\ast +} \overline D^0\pi^+} = 15.8(7)(3)\,,\nn
\eea 
where we symmetrized the second error bar. That value leads to the theoretical prediction of
\bea
\Gamma(D^{\ast +}\to \overline D^0 \pi^+)=\left(50\pm 5\pm 6\right)~\kev .
\eea
Our value is consistent with the result for $g_c$ (or $g_{D^{\ast +} \overline D^0\pi^+}$) extracted from the experimentally measured width of the charged vector meson, $\Gamma(D^{\ast \pm})$, after using the isospin symmetry to relate the decays to a charged and to a neutral pion in the final state. Furthermore, we were able to predict $\Gamma(D^{\ast 0})=(53\pm 5\pm 7)$~keV. 

The second error in all of the above results reflects our estimate of the uncertainty arising from the chiral extrapolation, i.e. the way one can relate the coupling $g_c$ directly computed on the lattice to the one corresponding to the physical pion mass. That error can be reduced in a new generation of lattice QCD simulations performed closer to the realistic physical situation, namely the lighter pions and by including the effect of $\nf=2+1+1$ dynamical quarks in the QCD gauge field configurations.

\section*{Acknowledgements}
We thank  the European Twisted Mass Collaboration for making their gauge field configurations publicly available, and in particular Silvano Simula for the discussions related to the subject of this paper. 
Computations of the relevant correlation functions are made on GENCI-CINES, under the Grant 2012-056806.

\end{document}